 \documentclass[nohyper,12pt,letterpaper]{JHEP3}
\usepackage{epsfig}
\usepackage{graphics}
\usepackage{float}
 
 \title{Gauge Coupling Unification in MSSM + 5 Flavors}

 \author{Jeff L.\,Jones\\
 Department of Physics and SCIPP\\
 University of California, Santa Cruz, CA 95064\\
 E-mail: \email{jeff@physics.ucsc.edu}}

\abstract{We investigate gauge coupling unification at 2-loops for theories with 5 extra vectorlike SU(5) fundamentals added to the MSSM.  This is a borderline case where unification is only predicted in certain regions of parameter space.  We establish a lower bound on the scale for the masses of the extra flavors, as a function of the sparticle masses.  Models far outside of the bound do not predict unification at all (but may be compatible with unification), and models outside but near the boundary cannot reliably claim to predict it with an accuracy comparable to the MSSM prediction.  Models inside the boundary can work just as well as the MSSM.}

 \keywords{Gauge Coupling Unification, RG Equations, Supersymmetry}
 \received{} \accepted{}
 \preprint{hep-th{08}\\\\scipp-2008/14 \\}
 
\begin{document}


\section{Introduction}
\label{introduction}

Low energy supersymmetry is a leading candidate for solving the hierarchy problem.  But its enduring popularity among model builders is in part due to another appealing feature which the simplest models of low energy supersymmetry possess:  in particular, the Minimal Supersymmetric extension of the Standard Model (MSSM) predicts gauge coupling unification.  Or phrased in another way, the combination of the MSSM and the assumption of grand unification together successfully predicted (roughly) the right experimentally measured value of $\alpha_3$ at the weak scale.

Despite this encouraging success, the MSSM by itself leaves many questions unanswered.  It contains many arbitrary parameters which must be set to certain values for purely phenomenological reasons.  However, it has proven difficult to construct more natural models of low energy SUSY breaking which still maintain coupling constant unification.  It is therefore of interest to identify which of the models that extend the MSSM can claim to maintain its prediction for $\alpha_3$ at a comparable level of accuracy.  This may help provide direction in the search for a fully plausible theory of supersymmetry breaking.

There are two primary ways in which a prediction for coupling constant unification can fail.  One problem is perturbativity.  If the coupling constants become too large before the would-be unification scale is reached, the perturbative expansion breaks down and the ability to reliably make predictions based on perturbation theory is lost.  The other problem is accuracy--given all coupling constants are still perturbative, do they meet at the same energy scale, or is there a mismatch?  Equivalently, how well do they predict the measured value of $\alpha_3$, given the assumption of unification?  Is the accuracy better or worse than the MSSM prediction?

We find that while the accuracy question depends sensitively on the details of the model, the answer to the perturbativity question is relatively robust and depends mostly on a few general features of the model.  The most important factor for perturbativity is the number of matter fields (with Standard Model gauge quantum numbers) added.  The more matter which appears in the loops, the less asymptotically free the theory becomes, causing the perturbative expansion to break down earlier as the gauge coupling constants are run up into the UV.

With 5 or more vectorlike SU(5) fundamentals\footnote{The reason only complete SU(5) multiplets are considered is because the accuracy of unification is usually destroyed, even at 1-loop, if incomplete multiplets are present.  There are some exceptions to this (\cite{Martin:1996zb},\cite{Calibbi:2009cp}).} added to the MSSM at the electroweak scale, the theory becomes strongly coupled before unification.  However, if the masses of the extra flavors are heavy enough, this statement does not apply.  For the case of exactly 5 extra flavors, the mass scale of the extra flavors becomes important for determining perturbativity.  This is the primary issue explored in this paper:  in the borderline case of 5 extra flavors, what are the perturbativity bounds on the masses of those flavors?  These bounds on fundamentals can also be translated into bounds on matter transforming in higher representations.  For example, each additional matter multiplet transforming in a 10 representation contributes the same to the RG equations as 3 additional 5 representations.

While some authors may choose to ignore models with 5 extra flavors simply because the perturbativity of the gauge couplings is questionable, there have been no careful studies to our knowledge of when this attitude is justified and why.  Our goal is to analyze the situation carefully and provide more justification for rejecting such models, when appropriate, and to identify the regime of parameter space where perturbativity is not a concern.

The question of which terms to keep in an asymptotic series when its expansion parameter is becoming strong is a somewhat ambiguous one.  We suggest a method for resolving the ambiguity and estimating the theoretical uncertainty involved in the prediction for $\alpha_3$ at the weak scale.  Using this method, we derive lower bounds on the masses of the extra flavors.

In Section~\ref{RGequations}, we review the relevant 1- and 2-loop renormalization group equations for the MSSM, and for the MSSM with extra vectorlike SU(5) fundamentals added.  In Section~\ref{thresholds}, we discuss the issue of 1-loop threshold corrections, both at the electroweak scale and at the GUT scale, which compete with 2-loop effects of the RG equations in determining the value of $\alpha_3$ at the electroweak scale.  In Section~\ref{methodology}, we describe our methodology for analyzing gauge coupling unification.  In Section~\ref{bounds}, we discuss the resulting perturbativity bounds on the scale of the extra flavors for various slices of parameter space.  The main result is a plot of a suggested lower bound on the extra flavors as a function of the scale of the superpartners assuming degenerate superpartner masses and a non-extremal value for $\tan\beta$.  Several modifications to this curve are then discussed, including heirarchies in the masses, large and small $\tan \beta$, and the addition of a coupling to an extra singlet field.  Some of these plots are included in this section, while others are saved for the appendix.  In Section~\ref{conclusions}, we draw general conclusions and make some closing remarks.

\section{RG Equations With Extra Flavors}
\label{RGequations}

At 2-loops, the RG equations for the gauge couplings are (\cite{Giudice:2004tc},\cite{Martin:1993zk}):

\begin{eqnarray}
\label{gaugeRGs}
(4\pi)^2 \frac{d}{dt} g_i = A_i g_i^3
    + \frac {g_i^3}{(4\pi)^2} \Big[\sum^3_{j=1} B_{ij} g_j^2 
    - \sum_{x=u,d,e} C_{i,x} {\rm Tr}(Y_x^{\dagger}Y_x)
\Big]
\end{eqnarray}
where $g_i$ are the Standard Model gauge couplings ($i=1,2,3$) with normalization based on SU(5) generators, $Y_x$ are the Yukawa coupling constants $Y_u$, $Y_d$, and $Y_e$ (each a $3\times3$ matrix for generations), and $t$ is the log of the renormalization scale.
\vskip 10pt
\noindent

For a supersymmetric theory with $N$ chiral matter superfields $\Phi_a$ ($a = 1\dots N$) in representations $R_{a,i}$ of gauge group $G_i$, the coefficients $A_i$ and $B_{ij}$ are given by \cite{Martin:1993zk}:
\begin{eqnarray}
A_i &=& \sum_a S(R_{a,i})-3C(G_i) \\
B_{ij} &=& -6[C(G_i)]^2 + 2C(G_i)\sum_a S(R_{a,i}) + 4\sum_a S(R_{a,i}) C(R_{a,j})]
\end{eqnarray}
where $C(R_{a,i})$ is the quadratic Casimir operator for the representation of the superfield $\Phi_a$ in the gauge group $G_i$, and $S(R_{a,i})$ is the corresponding Dynkin index.  For Yukawa couplings between the matter superfields $\Phi_a$ of the general form $\frac{1}{3!} Y^{abc}\Phi_a \Phi_b \Phi_c$, the term in the 2-loop RG equations from the Yukawas is:
\begin{equation}
-\sum_{a,b,c} Y^{abc} Y^{*abc} C(R_{c,i})/d(G_i) \label{generalYukawaContribution}
\end{equation}
The coefficients $C_{i,x}$ of the MSSM Yukawa couplings ${\rm Tr}(Y_x^{\dagger} Y_x)$ can then be extracted from this.

Plugging in MSSM values for these Casimirs and summing everything up leads to the following coefficients (\cite{Giudice:2004tc},\cite{Martin:1993zk},\cite{Ghilencea:1997mu},\cite{Bjorkman:1985mi}):

$$
A_i = \pmatrix{ \frac{33}{5}  \cr
                    1 \cr
                    -3 \cr
}
$$
\begin{eqnarray}
\label{MSSM}
B_{ij} = \pmatrix{ \frac{199}{25} & \frac{27}{5} & \frac{88}{5}  \cr
                    \frac{9}{5}   &  25         &   24      \cr
                    \frac{11}{5}  &  9          &   14       \cr
}
\end{eqnarray}
$$
C_{i,x} = \pmatrix{  \frac{26}{5}  & \frac{14}{5} & \frac{18}{5} \cr
                      6          &   6          &   2     \cr
                      4          &   4          &   0     \cr
}
$$
If n additional pairs of SU(5) 5 and $\overline{5}$ fundamentals are added, they become:

$$
A_i = \pmatrix{ \frac{33}{5}+n  \cr
                   1+n \cr
                   -3+n \cr} \\
$$
\begin{eqnarray}
\label{MSSMplus5}
B_{ij} = \pmatrix{ \frac{199}{25}+\frac{7}{15}n & \frac{27}{5}+\frac{9}{5}n & \frac{88}{5}+\frac{32}{15}n  \cr
                    \frac{9}{5}+\frac{3}{5}n  &  25+7n         &   24      \cr
                    \frac{11}{5}+\frac{4}{15}n  &  9          &   14+\frac{34}{3}n \cr
}
\end{eqnarray}
$$
C_{i,x} = \pmatrix{  \frac{26}{5}  & \frac{14}{5} & \frac{18}{5} \cr
                      6          &   6          &   2     \cr
                      4          &   4          &   0 \cr
}
$$
These agree with the formulas in \cite{Ghilencea:1997mu} for n=$\frac{1}{2} n_5$.
\vskip 10pt

The Yukawa couplings are also renormalized.  At 1-loop, the RG equations for general supersymmetric Yukawa couplings are\cite{Martin:1993zk}:

\begin{equation} \label{generalYukawaEvolution}
(4\pi)^2\frac{d}{dt} Y^{abc} = \frac{1}{2} Y^{abd} Y^{*dfg} Y^{cfg} - \sum_i 2 C(R_{c,i}) Y^{abc} g_i^2 + (c \leftrightarrow a) + (c \leftrightarrow b)
\end{equation}
For the Yukawa couplings of the MSSM, this becomes\cite{Martin:1993zk}:
\begin{eqnarray}
(4\pi)^2\frac{d}{dt} Y_u & = & Y_u \Big[ 3{\rm Tr}(Y_u Y_u^{\dagger})+3Y_u^{\dagger} Y_u + Y_d^{\dagger} Y_d - \frac{16}{3} g_3^2 - 3g_2^2 - \frac{13}{15} g_1^2 \Big] \\
(4\pi)^2\frac{d}{dt} Y_d & = & Y_d \Big[ {\rm Tr}(3 Y_d Y_d^{\dagger} + Y_e Y_e^{\dagger})+3Y_d^{\dagger} Y_d + Y_u^{\dagger} Y_u - \frac{16}{3} g_3^2 - 3g_2^2 - \frac{7}{15} g_1^2 \Big] \\
(4\pi)^2\frac{d}{dt} Y_e & = & Y_e \Big[ {\rm Tr}(3 Y_d Y_d^{\dagger} + Y_e Y_e^{\dagger})+3Y_e^{\dagger} Y_e -3g_2^2 - \frac{9}{5} g_1^2 \Big]
\end{eqnarray}

Because of the way in which the Yukawa couplings enter into the gauge coupling RG equations (they are always multiplied by at least a factor of $g^3$), the 1-loop Yukawa RG's given above are sufficient for a 2-loop calculation of the evolution of the gauge couplings.  For the purposes of this paper, we will be treating most of the entries in the Yukawa matrices as zero, only keeping the top and bottom Yukawa couplings.

$$
Y_u \approx \pmatrix{ 0 & 0 & 0 \cr
               0  & 0 & 0 \cr
               0  & 0   & y_t \cr
}
$$
$$
Y_d \approx \pmatrix{ 0 & 0 & 0 \cr
               0  & 0 & 0 \cr
               0  & 0   & y_b \cr
}
$$
$$
Y_e \approx \pmatrix{ 0 & 0 & 0 \cr
               0  & 0 & 0 \cr
               0  & 0   & 0 \cr
}
$$

\vskip 10pt

Another possible term which can arise in supersymmetric theories with extra fundamentals is a coupling $\lambda S M\overline{M}$ where $S$ is a singlet field (which can, for example, be used to help generate the $\mu$ term for the Higgs) and $M_i$ and $\overline{M}_i$ are the $n$ extra flavors ($i = 1 \dots n$).  In a gauge mediation context, $M$ and $\overline{M}$ are the messengers which transmit SUSY breaking to the visible sector.  If this term is present in the SU(5)-symmetric theory at the GUT scale, then there will be two such terms at energies where SU(5) is broken to ${\rm SU(3)\times SU(2)\times U(1)}$, $\lambda_2 S M^{(2)}_a \overline{M}^{(2)}_a$ and $\lambda_3 S M_b^{(3)} \overline{M}_b^{(3)}$ ($a = 1,2$, and $b = 1 \dots 3$).  The coupling constants $\lambda_2$ and $\lambda_3$ are equal at the GUT scale, but evolve separately in the broken phase below the GUT scale.

The 1-loop RG equations for the evolution of $\lambda_2$ and $\lambda_3$ can be derived from equation \ref{generalYukawaEvolution}.  The result is:

\begin{eqnarray}
(4\pi)^2 \frac{d}{dt} \lambda_2 & = & 4\lambda_2^3 + 3\lambda_2 \lambda_3^2 - 3g_2^2 \lambda_2 - \frac{3}{5}g_1^2\lambda_2 \\
(4\pi)^2 \frac{d}{dt} \lambda_3 & = & 5\lambda_3^3 + 2\lambda_3 \lambda_2^2 - \frac{16}{3}g_2^2 \lambda_3 - \frac{4}{15}g_1^2\lambda_3
\end{eqnarray}
The contribution of $\lambda_2$ and $\lambda_3$ to the RG equations for the gauge couplings can be derived from equation~\ref{generalYukawaContribution}:

\begin{eqnarray}
(4\pi)^2 \beta_{g_1} & = & \dots -6g_1^3\lambda_2^2 - 4g_1^3\lambda_3^2 \\
(4\pi)^2 \beta_{g_2} & = & \dots -10g_2^3\lambda_2^2 \\
(4\pi)^2 \beta_{g_3} & = & \dots -10g_3^3\lambda_3^2
\end{eqnarray}

\section{Threshold Effects}
\label{thresholds}

Given the assumption of unification, the predicted value of $\alpha_3(m_Z)$ can be written as a function of other weak-scale observables, by computing the values for $\alpha_1$ and $\alpha_2$ at the weak scale, evolving them up to the scale where they cross, and then evolving $\alpha_3$ back down from there.  To leading order in the coupling constants, this function can be computed simply by integrating the 1-loop RG equations.  However, at next order, the terms coming from integrating the 2-loop RG equations compete with terms arising from 1-loop threshold effects, some from the weak scale and others from the GUT scale.  Both of these effects are model dependent\cite{Bagger:1996ei}.

There are two types of 1-loop weak scale thresholds.  The first type are the leading logarithms, which arise from the RG equations themselves and the fact that the contribution from each mass comes in at a different scale.  Including the leading logarithms is an alternative to integrating a separate set of RG equations between every mass scale in the theory.  They allow one to include all MSSM particles in the RG equations starting at $m_Z$, even though many of the MSSM particles are heavier than $m_Z$.  These two approaches differ at 2-loop, but because all of the MSSM particles are close on a logarithmic scale to $m_Z$ compared to the full integration range (from $m_Z$ to $m_{GUT}$), the resulting difference in the GUT couplings as a function of the weak-scale parameters is comparable to 3-loop effects.  In addition to these logarithmic terms, there are also ``finite'' 1-loop radiative corrections to $\sin^2\hat{\theta}_{W}$, which affect the determination of $\hat{\alpha}_1(m_Z)$ and $\hat{\alpha}_2(m_Z)$ from electroweak observables.

The $\overline{MS}$ renormalization scheme does not preserve supersymmetry, so it is appropriate for us to use the $\overline{DR}$ scheme instead\cite{Martin:1993yx}.  All of the variables with hats in this section represent $\overline{DR}$ parameters.  The parameter $\sin^2\hat{\theta}_W$ can be defined by the relation (\cite{Pierce:1996zz},\cite{Bagger:1994ab}):

$$
\cos^2 \hat{\theta}_W \sin^2\hat{\theta}_W = \frac{\pi \hat{\alpha}(m_Z)}{\sqrt 2 m_Z^2 G_F (1-\Delta r)}
$$
where $G_F = 1.16637(1)\times 10^{-5}~{\rm GeV}^{-2}$ \cite{pdb2008} is the Fermi constant measured from muon decay, $\Delta r$ is a sum of a long list of radiative corrections, and $\hat{\alpha}(m_Z)$ is the value of the electromagnetic coupling at the weak scale which is computed from:

$$
\hat{\alpha}(m_Z) = \frac{\alpha_{EM}}{1-\Delta \hat{\alpha}}
$$
$\alpha_{EM} = 1/137.035999679(94)$ is the precisely measured value of the vacuum fine structure constant\cite{pdb2008}.  $\Delta \hat{\alpha}$ represents the radiative corrections to $\hat{\alpha}$.  Because the value of $\hat{\alpha}$ is used to compute $\sin^2\hat{\theta}_W$, $\Delta \hat{\alpha}$ also affects the value of $\sin^2\hat{\theta}_W$, indirectly.

At 1-loop, the radiative corrections to $\hat{\alpha}$ and $\hat{\alpha_3}$ in $\overline{DR}$ are given by\cite{Pierce:1996zz}:

\begin{eqnarray}
\Delta \hat{\alpha} & = & 0.0682 \pm 0.0007 - \frac{\alpha_{EM}}{2\pi}\Big\{-7\ln\Big(\frac{m_W}{m_Z}\Big) + \frac{16}{9}\ln\Big(\frac{m_t}{m_Z}\Big) \cr
& & + \frac{1}{3}\ln\Big(\frac{m_{H+}}{m_Z}\Big) + \frac{4}{9}\sum_{i=1}^{6} \ln\Big(\frac{m_{{\tilde u}_i}}{m_Z}\Big) + \frac{1}{9}\sum_{i=1}^{6} \ln\Big(\frac{m_{{\tilde d}_i}}{m_Z}\Big) \cr
& & + \frac{1}{3}\sum_{i=1}^{6} \ln\Big(\frac{m_{{\tilde e}_i}}{m_Z}\Big) + \frac{4}{3}\sum_{i=1}^{2}\ln\Big(\frac{m_{{\tilde \chi}^{+}_i}}{m_Z}\Big)\Big\} \cr
\Delta \hat{\alpha_3} & = & \frac{\alpha_3(m_Z)}{2\pi}\Big\{\frac{1}{2}-\frac{2}{3}\ln \Big(\frac{m_t}{m_Z}\Big)-2\ln\Big(\frac{m_{\tilde g}}{m_Z}\Big)-\frac{1}{6}\sum_{i=1}^{12} \ln\Big(\frac{m_{{\tilde q}_i}}{m_Z}\Big)\Big\} \cr
\hat{\alpha}_3(m_Z) &=& \frac{\alpha_3(m_Z)}{1-\Delta\hat{\alpha}_3}
\end{eqnarray}
where $\alpha_3(m_Z)$ on the righthand side of the equations represents the accepted $\overline{MS}$ value of the strong coupling constant at the weak scale.

The supersymmetric corrections to $\hat{\alpha}(m_Z)$ and $\hat{\alpha_3}(m_Z)$ depend in a simple way on only two masslike parameters, each a particular weighted geometric mean of different superpartner masses:

\begin{eqnarray}
M_{\alpha} & = & \Big[m_{H^+}^{1/3} \prod_{i=1}^{6} m_{{\tilde u}_i}^{4/9} \prod_{i=1}^{6} m_{{\tilde d}_i}^{1/9} \prod_{i=1}^{6} m_{{\tilde e}_i}^{1/3} \prod_{i=1}^2 m_{\chi^+_i}^{4/3}\Big]^{3/25} \\
M_{\alpha_3} & = & \Big[m_{\tilde g}^2 \prod_{i=1}^{12} m_{{\tilde q}_i}^{1/6}\Big]^{1/4}
\end{eqnarray}
The calculation of the $\sin^2\hat{\theta}_W$ corrections, by comparison, is far more complicated, involving pages of finite corrections rather than just logarithmic corrections, and depends on a large number of parameters.  (These expressions are hidden in the parameter $\Delta r$ above, and can be found in reference \cite{Pierce:1996zz}.)  It is therefore useful to consider $\sin^2\hat{\theta}_W$, $\tan\beta$, $M_{\alpha}$, and $M_{\alpha_3}$ as the inputs, rather than the entire MSSM parameter space.  In addition, to obtain even approximate gauge coupling unification, one must require a correlation between the two SUSY mass parameters, the scale of the extra flavors, and $\sin^2\hat{\theta}_W$.  This can be used to eliminate $\sin^2\hat{\theta}_W$ as an input, making it possible to represent the perturbativity bounds visually with a small number of plots.

\section{Methodology}
\label{methodology}
The 2-loop gauge coupling RG equations, coupled to the 1-loop Yukawa RG equations, were integrated numerically from the weak scale (defined as $m_Z$) up to the unification scale.  The unification scale, for our purposes, is defined as the scale $M_{GUT}$ at which $\alpha_1(M_{GUT}) = \alpha_2(M_{GUT})$.  The boundary conditions imposed at the weak scale are the weak-scale $\overline{DR}$ parameters, determined in terms of physical observables as well as $M_{\alpha}$, $M_{\alpha_3}$, $\tan\beta$, and $\sin^2\hat{\theta}_W$ which represent the MSSM parameters.  A fifth input, the energy scale of the extra 5 flavors, is not involved in the weak-scale boundary conditions but is important in determining the evolution of the coupling constants.

One measure of how well the couplings unify is how close $\alpha_3(M_{GUT})$ is to the value of the other two gauge couplings at the unification scale.  On some plots in this paper, this discrepancy is reported as a percentage difference.  This way of measuring the accuracy of unification is most useful when considering what unknown GUT thresholds would be necessary in order to obtain perfect unification.  If the required GUT thresholds are too large (more than a few percent), then the theory looks implausible from the standpoint of unification.

Another measure for how well the couplings unify is to use the unification scale determined by $\alpha_1$ and $\alpha_2$ to predict the value of $\alpha_3$ at the weak scale, comparing it to the accepted $\alpha_3(m_Z)$.  This way of measuring the discrepancy is more useful for assessing whether particular models can claim to have made a successful prediction (or at least a retrodiction).

Our method was to evolve the couplings up in energy from $m_Z$ using the MSSM RG equations (\ref{gaugeRGs},\ref{MSSM}), until reaching the energy scale of the 5 extra flavors.  At that scale, the RG equations including 5 additional flavors (\ref{gaugeRGs},\ref{MSSMplus5} for $n=5$) were used to continue evolving the couplings into the UV.  Both 1-loop and 2-loop expressions for the running coupling constants were evolved simultaneously in this manor.  Depending on the input parameters, several different outcomes are possible during the numerical evolution.

One possibility is that even in the 1-loop approximation, one or more of the couplings becomes strong before unification ($\alpha_1 = \alpha_2$) is reached.  Obviously, in this case, there is no prediction for unification.  Another possibility is that the 1-loop couplings remain weak and unify, but one or more of the 2-loop couplings becomes strong (meaning $\alpha_i > 1$).  A third possibility is that both of the 1- and 2- loop couplings remain formally less than 1, but that a more restrictive condition on the couplings is violated before unification.  Above the scale where the extra matter lies, as the energy scale increases the gauge couplings all become stronger.  At some point before a coupling reaches $\alpha_i = 1$, the 2-loop contribution to the beta function begins to dominate the 1-loop contribution.  (As mentioned by \cite{Ghilencea:1997mu} and confirmed in our results, a typical place for this to happen is around $\alpha_i = 0.3$).  After this point, even though the couplings are still ``perturbative'' in a sense and the 1-loop beta function may be approximately useful, the 2-loop beta function is completely untrustworthy.

To make the reasoning here clearer, consider any asymptotic expansion in some parameter $\alpha$ about $\alpha=0$. Only a finite number of terms $n$ in the expansion converge on the true non-perturbative value of the function--while the rest of the terms beginning with $n+1$ diverge (as more terms are kept, they get further and further from the right answer, rather than closer to it).  As $\alpha$ becomes larger, the number of convergent terms $n$ decreases.  Although it depends on the function being expanded and there are known mathematical exceptions, a general rule of thumb for determining how many terms $n$ to keep is to look for the place where the $(n+1)^{th}$ term is larger than the $n^{th}$ term.  When the full non-perturbative function is unknown, this rule of thumb is the only guide we have for knowing how many terms in the series to keep.  Hence if the second term in the asymptotic expansion for the beta function is larger than the first term, it is unreasonable to expect that the second term adds any additional information to the first.  It is possible in some cases that it makes the approximation better, but generally it would make the approximation worse.  For example, in \cite{Ferreira:1996az} a case of the coupling constant becoming strong is presented where the beta function is argued to be Pade-Borel summable--under such an assumption the full beta function is computed and found to be closer to the 1-loop approximation than the 2-loop approximation.

Keeping this rule of thumb for asymptotic series in mind, more and more of the terms in the series become useless as the coupling constants become stronger.  The fewer the terms in the series we keep, the more theoretical uncertainty in the beta function and any predictions extracted from it.  After all of the terms have become unreliable, the coupling constant finally becomes ``strong,'' and the entire perturbation series has broken down.  Looking for a formal Landau pole in the RG equations overestimates the scale at which the series breaks down, so instead we have looked at which terms in the beta function are dominant and used that to estimate the theoretical uncertainty involved.

As will be discussed in Section~\ref{bounds}, the MSSM 2-loop prediction for $\alpha_3(m_Z)$ (which depends on the masses of the superpartners) is generally worse than the 1-loop prediction.  So it would appear that only 1-loop perturbativity is required in order to compete with the ``success'' of the MSSM prediction.  However, we argue that a stronger requirement--that the expansion at 2-loops is still reliable in the sense discussed above--is a more appropriate (though admittedly still somewhat arbitrary) condition for determining whether a model can compete with the MSSM in claiming to predict $\alpha_3(m_Z)$.  This was not an assumption going in to this project, but a conclusion that emerged after running numerical simulations of the evolution of the coupling constants for many different input parameters and comparing their fates.

A key issue here is estimating how reliable a 1-loop calculation is in the case where the 2-loop expansion breaks down before unification.  Already above, we have divided the parameter space into 3 regions:  one where the 2-loop expansion holds all the way to unification, another where the 1-loop holds but the 2-loop breaks down, and a third where the coupling becomes fully non-perturbative before unification and even the 1-loop calculation is meaningless.  Now we divide the second of these 3 regions into two subregions, by identifying a window where the 2-loop calculation fails near enough the unification scale that there is a way to estimate roughly the theoretical uncertainty involved in the 1-loop calculation.

Our method for estimating the theoretical uncertainty in the 1-loop computation goes as follows.  First, run the couplings up until the 2-loop term in the beta function for one of them (typically $\alpha_3$) becomes larger than the corresponding 1-loop term.  Above this point, we presume that the 1-loop approximation to that beta function is a better approximation than the 2-loop approximation (following the rule of thumb described above).  After that, continue to run the couplings up into the UV, but using the 1-loop beta function for the coupling which has become too strong for the 2-loop to make sense.  In the event that the same thing happens to another coupling, switch over to the 1-loop approximation for that beta function as well.  Assuming that unification is reached before any of the couplings become fully non-perturbative, use the value of $\alpha_1(M_{GUT}) = \alpha_2(M_{GUT})$ computed in this way to run all of the couplings back down to the weak scale.  On the way down, the same procedure is used where either the 1-loop or the 2-loop beta function for each coupling constant is used, depending on which is more appropriate.  Then compare this hybrid method for computing $\alpha_3(m_Z)$, which could perhaps be described as a ``1.5-loop calculation'', to the purely 1-loop calculation of $\alpha_3(m_Z)$ using the same input parameters.  The theoretical uncertainty in the 1-loop result is then estimated to be the difference between these two estimates for $\alpha_3(m_Z)$.

Another way to view the method of 1.5-loop calculation described above is as the continuum limit of a function which depends on the sum of a large discrete number of terms, each involving a different coupling, where some of the couplings are too strong to include their 2-loop contribution and others are weak enough to include their 2-loop contribution.  However, this ``sum'' is a bit exotic in that each term depends on the last, and the renormalization scale is different for each term.  Since the running is computed numerically by discretizing the integral, this is not only an analogy, but also a description of how the calculation was performed.

Our method has some arbitrariness, but it is only employed to get a rough idea of how trustworthy the 1-loop calculation is, for the purpose of deciding whether 1-loop perturbativity is enough to say that a given model predicts unification.  Based on this method, it appears that 1-loop perturbativity is not enough because of the large theoretical uncertainty when the coupling constant is becoming strong.

\section{Bounds}
\label{bounds}

The value of $\alpha_3(m_Z)$ has been measured by a number of different experiments.  The current global average is reported as $0.1176 \pm 0.002$ by the particle data group \cite{pdb2008}.  A global fit of all Standard Model parameters yields $\alpha_3(m_Z) = 0.1217 \pm 0.0017$, which disagrees by about $3.5\%$, more than $1\sigma$, from the average taking into account all experiments (including lattice QCD).  The disagreement could be statistical in nature, or it could be that adding physics beyond the Standard Model will reconcile these two values.  Non-supersymmetric GUT theories predict $\alpha_3(m_Z) = 0.073 \pm 0.001$, far too low.  The MSSM embedded in a minimal SU(5) GUT theory predicts $\alpha_3(m_Z) = 0.130 \pm 0.01$ at 2-loops.  This value is larger than the experimental value of $0.1176$ by $10.5\% \pm 8.5\%$.  This is far worse than the 1-loop estimate, but the large error bars on it (which are primarily due to the unknown GUT threshold effects) could bring it back into agreement.  Given that the MSSM only predicts the correct value of $\alpha_3$ to within $\sim 10\%$, we will not require any better accuracy from extensions of the MSSM.  The 1-loop MSSM prediction of $\alpha_3(m_Z)$ is $0.1178$, which is only $0.2\%$ larger than the experimental value.  However, since the 2-loop RG equations and the 1-loop thresholds make the prediction worse, the seemingly impressive closeness of the 1-loop prediction to the experimental value can only be regarded as a coincidence.

\begin{figure}[htb!]
\centering
\includegraphics[width=1.1\textwidth]{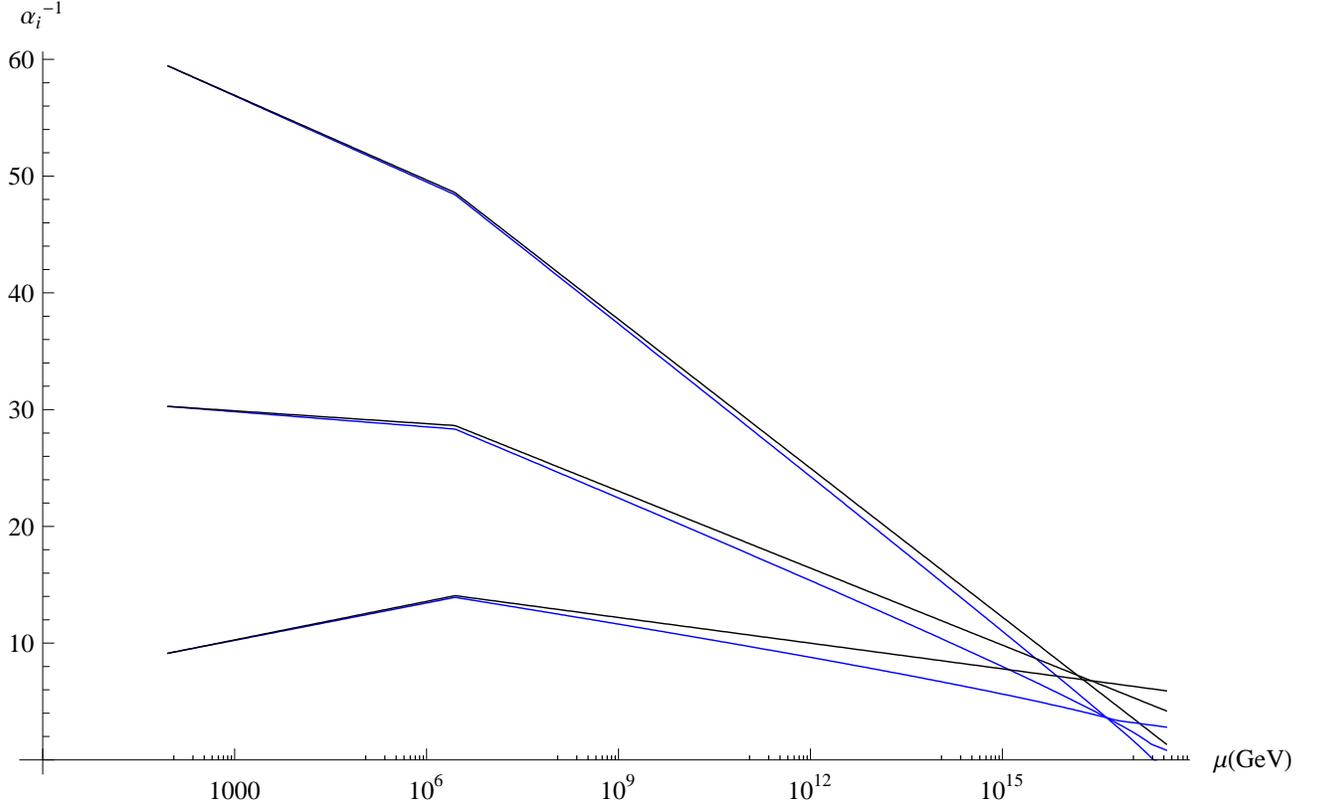}
\caption{1-loop and ``1.5-loop'' evolution of gauge couplings for SUSY partners at 250 GeV, 5 extra flavors at 2800 TeV, and $\sin^2\hat{\theta}_W = 0.2341$.  The coupling constants unify just before the 2-loop expansion is lost.}
\label{fig:hybridplot02}
\end{figure}

\begin{figure}[htb!]
\centering
\includegraphics[width=1.1\textwidth]{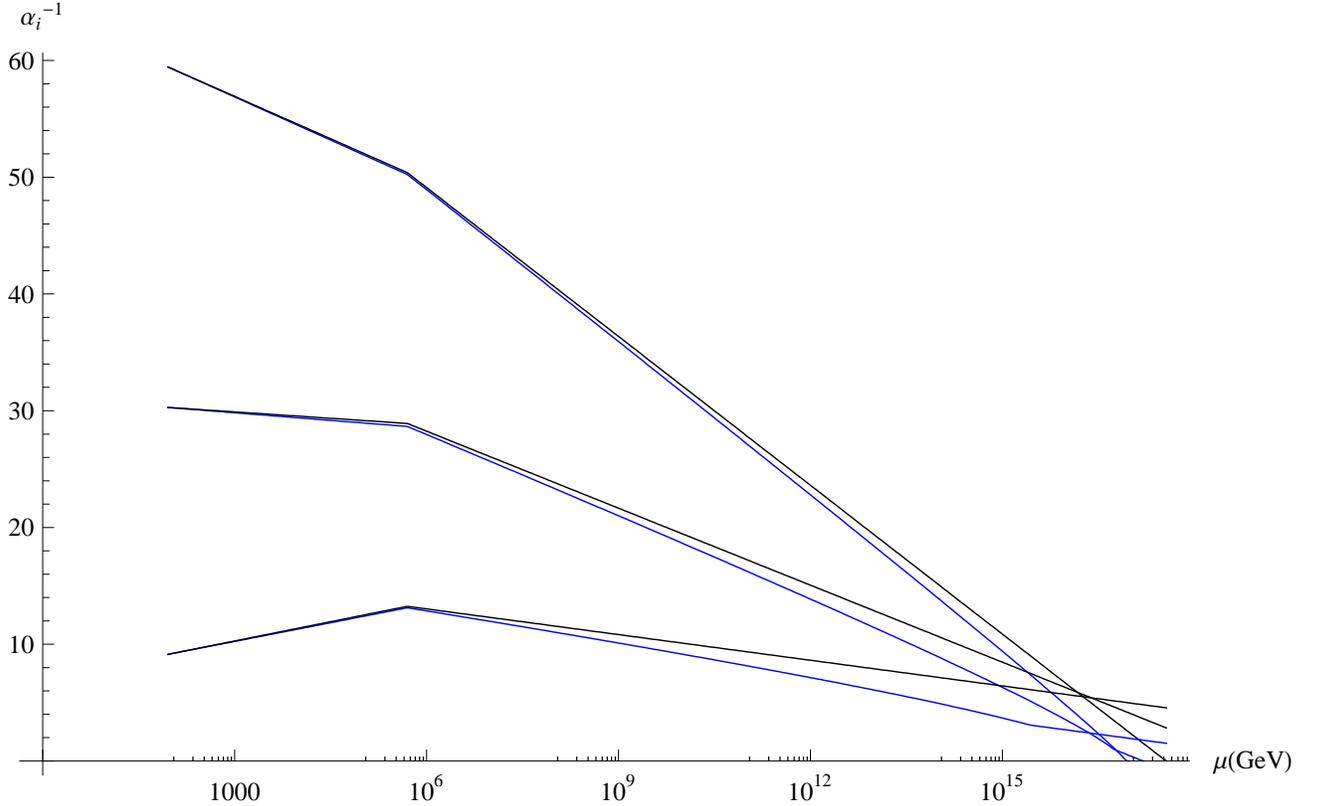}
\caption{1-loop and ``1.5-loop'' evolution of gauge couplings for SUSY partners at 250 GeV, 5 extra flavors at 500 TeV, and $\sin^2\hat{\theta}_W = 0.2341$.  The couplings do not unify before the 2-loop expansion is lost, and the 1.5-loop continuation does not unify with much accuracy.}
\label{fig:hybridplot01}
\end{figure}

Figures \ref{fig:hybridplot02} and \ref{fig:hybridplot01} show two examples of the procedure described in Section~\ref{methodology}:  the coupling constants are evolved using both 1- and 2-loop expansions simultaneously, and the 2-loop evolution switches over to 1-loop once the first and second terms in the corresponding beta function become of equal magnitude.  The black lines represent the 1-loop evolution, while the blue lines represent the ``1.5-loop'' evolution (2-loop or 1-loop, whichever is more appropriate as the energy scale changes).  In both plots, there is a kink in the evolution at the scale where the 5 extra flavors appear.  There are also kinks where the 2-loop evolution switches to 1-loop for $\alpha_3$ and $\alpha_2$.  In Figure~\ref{fig:hybridplot02}, the extra matter is at 2800 TeV, and unification at 2-loops happens just before the 2-loop expansion is lost.  In Figure~\ref{fig:hybridplot01}, the extra matter is at a lower scale of 500 TeV, and the 2-loop becomes bad before unification happens.  Figure~\ref{fig:hybridplot01} is an example of the window in parameter space nearby where 2-loop perturbativity fails where the theoretical uncertainty in the 1-loop calculation can be roughly estimated.  As is typical, the accuracy of the 1.5-loop unification is poor and when the couplings are run back down to the weak scale, there is a fairly large uncertainty in the prediction for $\alpha_3(m_Z)$.  For this case, the 1-loop prediction for $\alpha_3$ is 0.1136 while the 1.5-loop prediction for $\alpha_3$ is 0.1394, for a theoretical uncertainty of about 23\% (one is too large and the other is too small).  This uncertainty typically becomes worse as the energy scale for the extra flavors is lowered further, becoming incalculable once the 1.5-loop calculation becomes non-perturbative before unification.  As is evident from the plot, the uncertainty in the strength of the unified coupling at the GUT scale is even worse, since the 1- and 1.5-loop crossing points are far apart from each other.

The value of $\sin^2\hat{\theta}_W(m_Z)$ used here was chosen for these plots simply because it provides good unification when the threshold for the extra matter is high enough.  The actual value depends on all of the SUSY masses and could take on many different values for fixed values of $M_\alpha$ and $M_{\alpha_3}$ (which are set equal in these plots).  While the Standard Model $\overline{MS}$ value of this parameter is known\cite{pdb2008} to be $0.23119\pm 0.00014$, the $\overline{DR}$ value in the MSSM can range all the way from 0.230 to 0.238 depending on the details of the SUSY spectrum\cite{Chankowski:1994ua}.

\begin{figure}[htb!]
\centering
\includegraphics[width=1.1\textwidth]{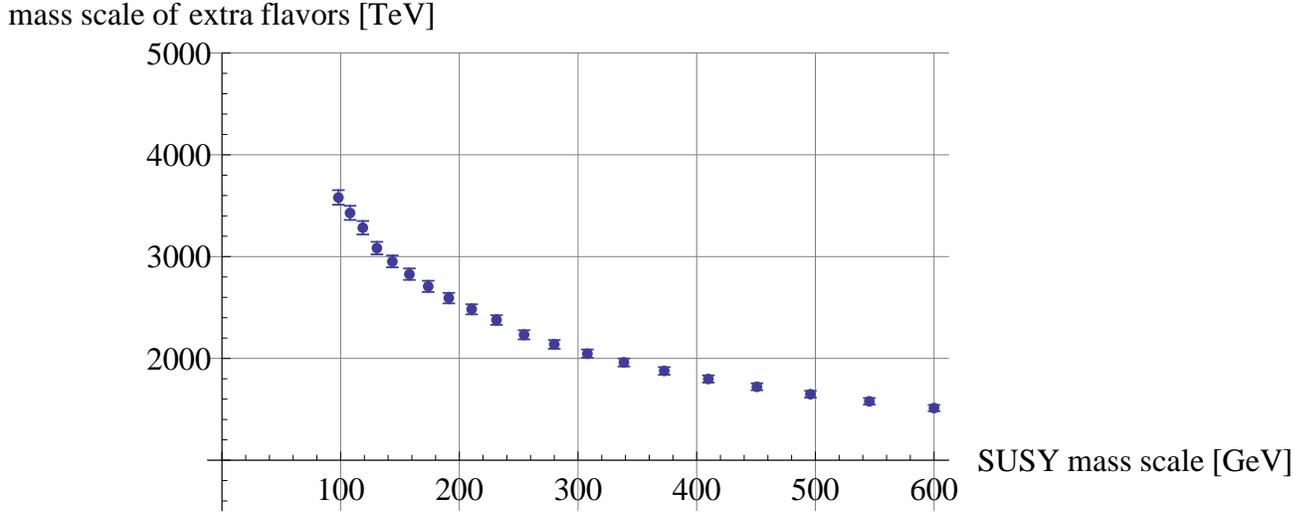}
\caption{A plot of the boundary below which 2-loop perturbativity is lost before unification occurs. $\sin^2\theta = .234$.  Error bars indicate vertical scanning resolution in numerical algorithm (2\%).}
\label{fig:threshplot07}
\end{figure}

\begin{figure}[htb!]
\centering
\includegraphics[width=1.1\textwidth]{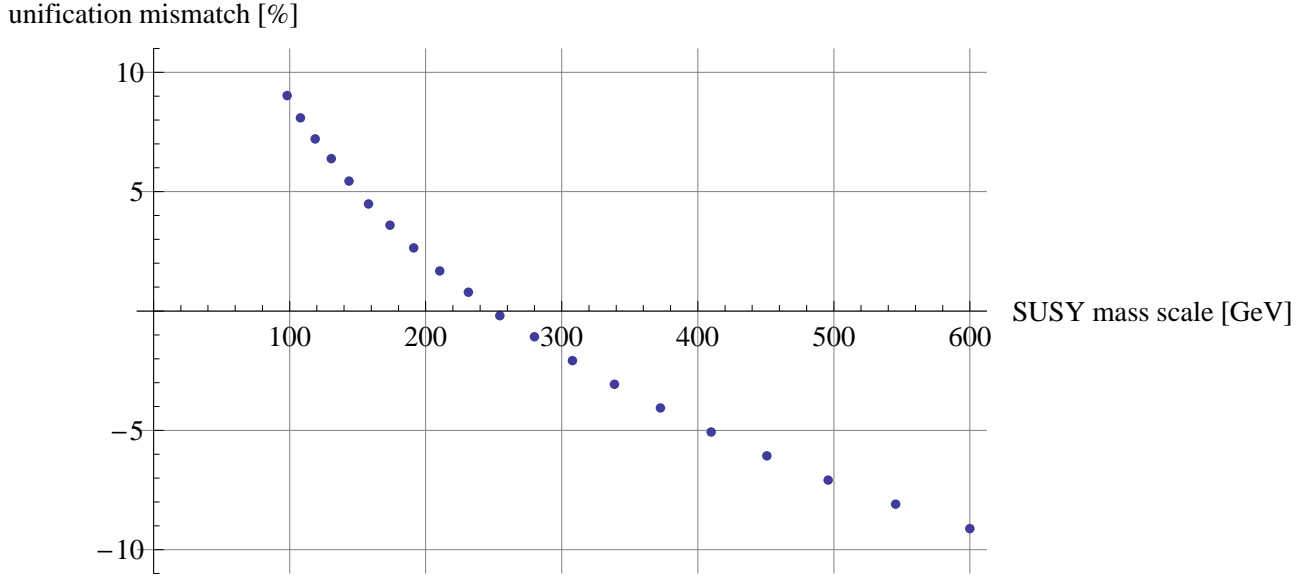}
\caption{A plot of the magnitude of the GUT threshold correction required in order to make the gauge couplings unify, for $\sin^2\hat{\theta}_W$ = 0.234 where the 2-loop expansion is barely perturbative at unification.}
\label{fig:errorplot07}
\end{figure}

After experimenting with different input parameters, it appears that the most important boundary in parameter space is between the region where 2-loop perturbativity holds all the way to unification, and where the 2-loop expansion in at least one of the couplings breaks down before unification.  Figure~\ref{fig:threshplot07} shows a plot of this boundary for a fixed value of $\sin^2\hat{\theta}_W = 0.234$, and for a single scale for the superpartners of $M_{SUSY} = M_{\alpha} = M_{\alpha_3}$.  This was computed by scanning over the corresponding slice of parameter space and running the couplings from weak to GUT scale many times for each point on the plot.  However, this plot is not the most useful because the accuracy of unification is sensitive to $\sin^2\hat{\theta}_W$, and different values of $\sin^2\hat{\theta}_W$ are required for different values of $M_{SUSY}$.  Figure~\ref{fig:errorplot07} shows a plot of the percentage difference between the coupling constants at the GUT scale $\Big(\frac{\alpha_3(M_{GUT})-\alpha_1(M_{GUT})}{\alpha_1(M_{GUT})} \times 100\%\Big)$.  Equivalently, this is an estimate of the magnitude of the threshold corrections required in the GUT theory to attain perfect unification.  For this value of $\sin^2\hat{\theta}_W$, the scale of the SUSY partners must be somewhere around 220~GeV in order to avoid requiring large GUT threshold corrections.



%

\begin{figure}[htb!]
\centering
\includegraphics[width=1.1\textwidth]{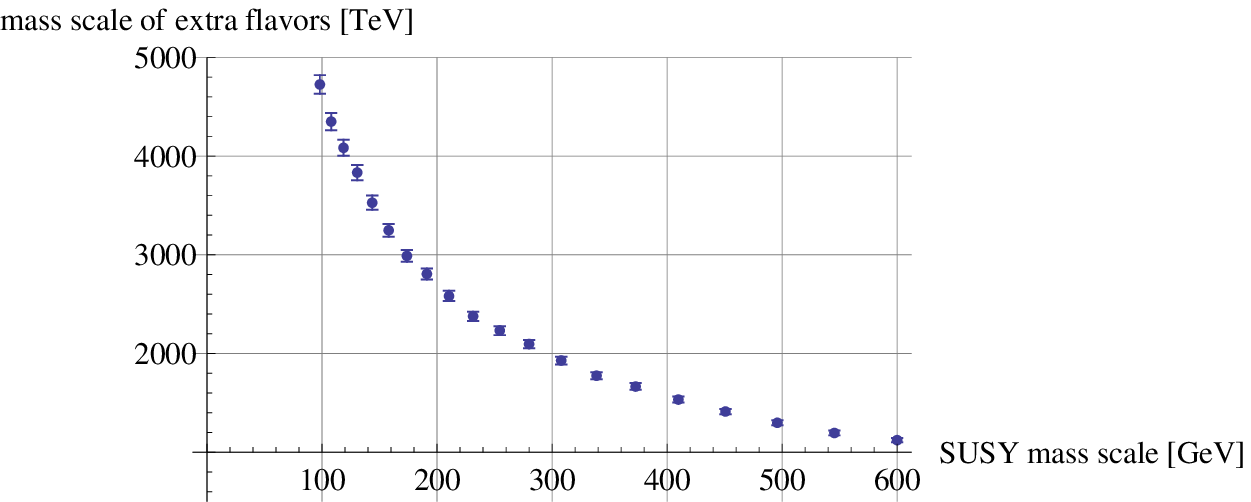}
\caption{A plot of the boundary below which 2-loop perturbativity is lost before unification occurs, with $\sin^2\hat{\theta}_W$ tuned at each point to minimize unification error.  Linear-linear scale.  Error bars indicate vertical scanning resolution in numerical algorithm (2\%).}
\label{fig:threshplot01}
\end{figure}

To compensate for the problem of a fixed value of $\sin^2\hat{\theta}_W$ only allowing a narrow range of realistic values of $M_{SUSY}$ by realistic accuracy constraints, we have fed this curve back into another algorithm which finds the most appropriate value of $\sin^2\hat{\theta}_W$ to use for each point on the boundary.  This is done by scanning over different values of $\sin^2\hat{\theta}_W$ and finding the minimum unification error, rerunning the boundary-finding algorithm with the new array of $\sin^2\hat{\theta}_W$ values, and iterating this procedure until the new boundary no longer moves appreciably.  The result is the far more useful plot shown in Figure~\ref{fig:threshplot01}.

Figure~\ref{fig:threshplot01} is the central result of this paper.  It is a plot of the boundary where 2-loop perturbativity is lost, but where $\sin^2\hat{\theta}_W$ is tuned at each point to a value which maximizes the ability of the couplings to unify well.  With a few exceptions to be soon discussed, any model with 5 extra fundamentals whose masses fall below this boundary must either sacrifice accuracy or perturbativity (predictivity), regardless of what the value of $\sin^2\hat{\theta}_W$ is.  The exceptions to this are if the two SUSY scales set equal here are split, or if there is a Yukawa coupling which is very near its own perturbativity limit.  These exceptions can shift the bound somewhat, but not by much.  Notice that the lower bound in Figure~\ref{fig:threshplot01} is raised somewhat relative to the plot in Figure~\ref{fig:threshplot07} for a low SUSY scale, and relaxed somewhat for a higher SUSY scale.  None of the boundary points differ by more than a factor of 2.  The range of $\sin^2\hat{\theta}_W$ values required to generate this plot was from 0.2323-0.2356.

\begin{figure}[htb!]
\centering
\includegraphics[width=1.1\textwidth]{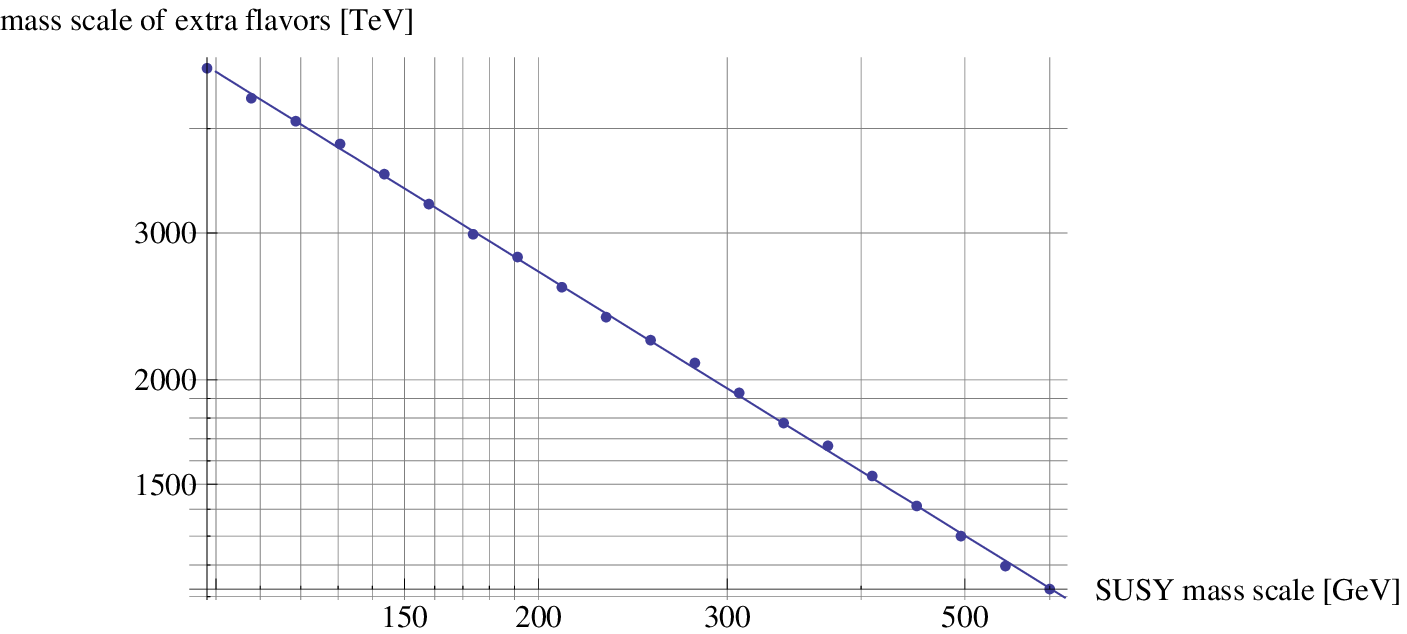}
\caption{A plot of the boundary below which 2-loop perturbativity is lost before unification occurs, with $\sin^2\hat{\theta}_W$ tuned at each point to minimize unification error.  Log-log scale.}
\label{fig:threshplotloglog01}
\end{figure}

On a log-log scale, the same plotted points can be closely approximated by a straight line (Figure~\ref{fig:threshplotloglog01}).
The empirical formula for the mass relationship at the boundary extracted from this is
$$
M_{5 {\rm flav}}^{0.557} M_{SUSY}^{1-0.557} \geq 40.0~{\rm TeV}
$$

In the above plots, the two relevant SUSY mass scales, $M_{\alpha}$ and $M_{\alpha_3}$, were taken to be equal.  However, there is no reason to assume there could not be a separation between them.  $M_{\alpha}$ is a weighted average of SUSY partners with electromagnetic charge, while $M_{\alpha_3}$ is a weighted average of SUSY partners with color charge.  If the squarks and gluinos are much heavier than the sleptons and charginos, then $M_{\alpha}$ and $M_{\alpha_3}$ would be separated.  In gauge mediation, there is a natural heirarchy between the color charged and color neutral sparticles.  However, because the squarks are included in both of the averages, the separation between $M_{\alpha}$ and $M_{\alpha_3}$ will not be as much as the heirarchy between the color charged and color neutral partner masses themselves.  For example, the typical heirarchy of between 3:1 and 6:1 in gauge mediation would only lead to a ratio of $M_{\alpha_3}/M_{\alpha} \approx 2$.

\begin{figure}[htb!]
\centering
\includegraphics[width=1.1\textwidth]{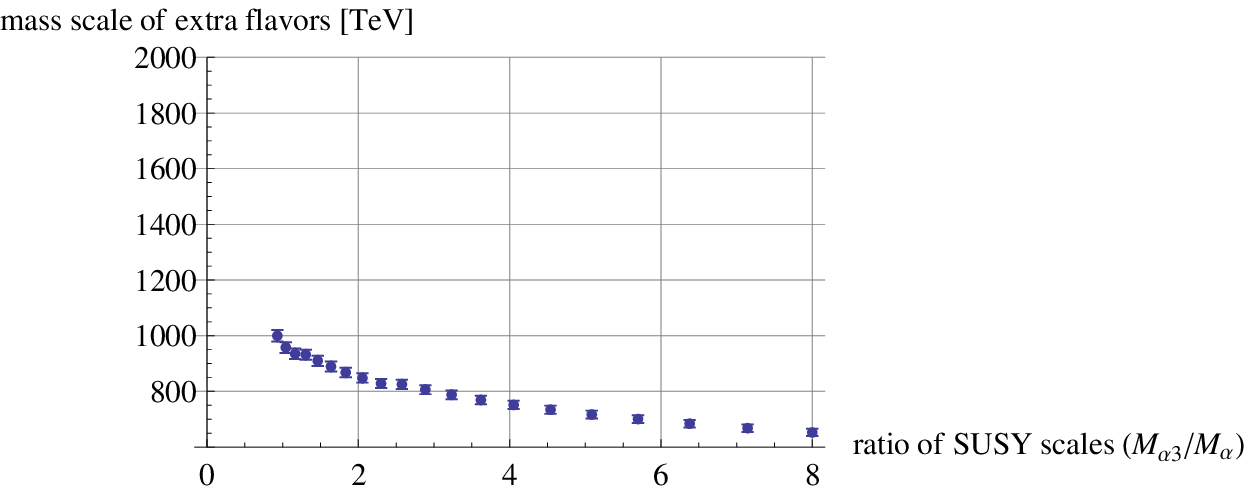}
\caption{A plot of the perturbativity boundary for $M_{\alpha_3} = 800~{\rm GeV}$, $M_{\alpha_3}/M_{\alpha} = 1 - 8$, and $\sin^2\hat{\theta}_W$ tuned to minimize unification error.}
\label{fig:threshplot06}
\end{figure}

After varying the scales independently, we found that most of the variation in the bound on the mass scale for the extra flavors is due to $M_{\alpha_3}$.  By comparison, the bound is relatively insensitive to changes in $M_{\alpha}$, but it has a weak effect which goes in the {\it opposite} direction as $M_{\alpha_3}$.  The result is that the bound is relaxed somewhat compared to $M_{\alpha} = M_{\alpha_3}$ if the ratio $M_{\alpha_3}/M_{\alpha}$ is large.  The case which allows the lowest scale for the extra flavors then, is when $M_{\alpha_3}$ is as high as possible and $M_{\alpha}$ is as low as possible.  Fig~\ref{fig:threshplot06} shows a plot of the perturbativity bound as a function of $M_{\alpha_3}/M_{\alpha}$ for fixed $M_{\alpha_3} = 800~{\rm GeV}$.  The required value of $\sin^2 \hat{\theta}_W$ for this range of parameter space is $0.2356-0.2367$.  A gauge mediation scenario with the squarks and gluinos at 800~GeV and the rest of the sparticles between 100 and 300~GeV would be represented by the $M_{\alpha_3}/M_{\alpha} = 2$ point on the plot ($M_{5 \rm flav} \ge 820$~TeV).

In order to achieve $M_{\alpha_3} = 800~{\rm GeV}$ and $M_{\alpha} = 100~{\rm GeV}$ (the value 8 on the independent axis of Fig~\ref{fig:threshplot06}), one would need the gluinos to be at 6.4~TeV and for the rest of the sparticles including the squarks to all be at 100~GeV.  Raising the squarks up closer to the gluinos would raise both $M_{\alpha}$ and $M_{\alpha_3}$.  The ratio between the scales would then be less than 8, however the overall bound would be slightly more relaxed.  In principle, it is possible to lower the lower bound on the extra 5 flavors all the way down to about 300~TeV, but only at the expense of requiring a large heirarchy and gluinos several hundreds of times heavier than the lightest supersymmetric partners.  This is much larger than could be attained naturally in gauge mediation (or any other SUSY breaking scenario we are aware of).

One may wonder why we have assumed the 5 extra flavors are mass degenerate, and if anything changes if they are non-degenerate.  If they are non-degenerate then the mass scale used here can be taken to be the geometric mean of all of the masses.  Because all of the extra flavors contribute the same type of term to the RG equations (each with a different mass), the terms can be combined in the same way we have combined the masses of the SUSY particles into two composite mass parameters.  This should hold at 2-loops as long as the masses are roughly near the same scale (within an order of magntiude or so).  The reason for this is the same reason that only 1-loop threshold corrections are required for a 2-loop unification analysis:  the distance on a logarithmic scale over which the RG equations differ is small compared to the entire range over which the couplings are run, and higher order corrections would matter only at higher loops.  So the answer here is that our analysis is valid as long as all 5 flavors are roughly on the same order of magntiude; but the situation may be different if there is a huge hierarchy in their masses.  We have nothing more to say about such models--it would be simple to extend our analysis to any single one, but it does not seem worth categorizing or testing all of the many possibilites at present.

In all of the above, we have set $\tan\beta = 10$.  However, the curves have been recomputed for other values of $\tan\beta$, and it was found that the resulting plots do not change noticeably at all unless $\tan\beta$ is close to its own perturbativity limit.  So the above plots can be taken to represent a wide range of values for $\tan\beta$, leaving out only a tiny bit of parameter space near the small $\tan\beta$ limit.

For large $\tan\beta$ there is a minimum value for the SUSY scale, in order for the bottom Yukawa coupling to stay perturbative until the unification scale.  If the SUSY scale is too low, then regardless of the masses of the extra flavors, the 2-loop calculation is lost before unification.  If the SUSY scale is high enough, then the lower bound on the mass scale for the extra flavors roughly matches the one for $\tan \beta = 10$.  The curve is very slightly lowered, but the difference is hardly noticeable on the plot and only slightly greater than the uncertainty inherent in the algorithm used.

A similar thing happens for small $\tan\beta$ ($\tan\beta \sim 1$), except in that case the top Yukawa is the important factor and the curve is shifted down noticeably from the non-extremal $\tan\beta$ result.  (The difference is because the top Yukawa has a larger effect on the gauge couplings through its larger coefficient in the RG equations.)  The bound is still within a factor of 2 of the non-extremal $\tan\beta$ bound, probably not significant enough of a difference for model builders to worry about.

Some plots of the extreme $\tan(\beta)$ bounds are included in the Appendix (Fig~\ref{fig:threshplot03},Fig~\ref{fig:threshplot04}).  The quoted values for $\tan\beta$ in the captions should not be taken too seriously, because the 2-loop Yukawa RG equations were not used, nor were the 1-loop Yukawa thresholds considered.  This leaves the value of $\tan(\beta)$ that corresponds to particular values of the $\overline{DR}$ Yukawa couplings approximate, however this uncertainty does not extend to their effect on the gauge couplings.  For our purposes (analyzing the perturbativity of the gauge couplings when the Yukawa couplings take extreme values), this should not matter.

The final modification we considered was the addition of a coupling between the extra fundamentals and an extra singlet field.  The single coupling $\lambda$ at the GUT scale runs as two separate couplings $\lambda^{(2)}$ and $\lambda^{(3)}$ below the GUT scale.  If they are unified at the GUT scale, the typical result is that $\lambda^{(3)}$ is greater than $\lambda^{(2)}$ after they are run down to the weak scale.  For most values $\lambda$ could take at the GUT scale, there is no discernable effect on the runnings of the gauge couplings.  But as with the top and bottom Yukawa couplings, the exception is when $\lambda$ is right near its own perturbativity limit ($\lambda = \sqrt{4\pi}$).  Fig~\ref{fig:lambdaplot01} in the appendix shows an example of this case and what happens when it is run down to the weak scale.

The result is that if $\lambda$ is near its perturbativity limit, then it can relax the bound in Fig~\ref{fig:threshplot01} by just barely more than a factor of 2.  Its effect can be larger than that of the top or bottom Yukawa couplings, because the coefficient in front of its term in the gauge coupling RG equations is larger.  In the appendix, Fig~\ref{fig:threshplot08} and Fig~\ref{fig:threshplot09} show two examples of how the 2-loop perturbativity boundary for the gauge couplings is shifted for fixed large values of $\lambda^{(2)}$ and $\lambda^{(3)}$ at the weak scale.  If they are too large, then there is an upper limit on the SUSY scale otherwise one or both of them will become non-perturbative before unification.  Ideally, it would be better to have plots based on a fixed value of $\lambda$ at the GUT scale rather than the weak scale.  The running depends on the SUSY scale as well as the scale for the extra flavors, and for much of the range plotted the two couplings do not actually unify (although they were chosen to come close for the middle of the range).  To do the analysis for fixed GUT scale coupling would be more complicated and could be a direction for future improvements, however our opinion is that it is unlikely to change the results significantly.

\section{\bf Conclusions}
\label{conclusions}

Dealing with asymptotic expansions near the limit where they are becoming non-perturbative is a murky business, and it is difficult to make precise or absolute claims.  Nevertheless, we hope we have convincingly demonstrated in this paper using a fairly reasonable methodology that models including 5 extra flavors below a certain scale in the neighborhood of a thousand TeV, depending on the masses of the superpartners, cannot reliably claim to make predictions of gauge coupling unification on par with the MSSM.  We have provided a fairly robust lower bound as a guideline for model builders to stay above, if they wish to do as well as the MSSM on this front.

Models above the bound can do just as well as the MSSM in predicting unification.  However, whether they actually predict unification (that is, whether the coupling constants meet each other or miss each other) depends on the details of the model--for instance, the wrong distribution of SUSY partners may lead to a value of $\sin^2(\theta_W)$ which predicts that the couplings do not unify.  From the point of view of gauge coupling unification, models with 5 extra flavors which predict unification should be considered just as seriously as candidates with fewer flavors.  But because the scale must be separated by at least a few orders of magnitude from the electroweak scale, the fine-tuning problems in these models will likely be more severe, and the prospects for detectable signatures at LHC worse.  Models below the bound can still be considered as candidates to solve other problems in high energy physics, but if they are to be believed then the successful MSSM prediction for $\alpha_3$ must be seen as a coincidence.

One example of a valid UV-completion of the Standard Model involving 5 or more extra flavors below the bound would be a duality cascade, where the couplings become strong at some energy scale and above that scale there is a dual description of the theory which is weakly coupled.  However, it is hard to know how to relate the couplings of the high energy theory to the couplings of the low energy theory.  Unless this challenge is overcome, these models cannot claim to predict unification either.

If the masses of the gluinos (and optionally, the squarks) were a couple orders of magnitude above the masses of the other sparticles, then the separation of scales could allow the 5 extra flavors to be lowered to a few hundred GeV, however the required heirarchy would be much larger than gauge mediation or other natural models of SUSY breaking would predict, and some of the motivation for low energy supersymmetry (ie, solving the heirarchy problem) would be undone.

Small $\tan\beta$ (and to a lesser extent large $\tan\beta$) can relax the perturbativity bounds slightly, but do not significantly change the other conclusions.  To say something more precise about particular values of $\tan\beta$ near 1, a more detailed analysis including 2-loop Yukawa coupling RGE's is needed.  Adding a coupling between the extra fundamentals and an extra singlet field has no effect unless the coupling is near its own perturbativity bound, in which case the effect is similar to the effect of the top Yukawa (and slightly larger).  If both of these effects were combined in just the right way, along with a heirarchy between color charged and color neutral objects, it is reasonable to believe that the bound in this paper could be relaxed even more.  However, to get such a model where all of these things line up favorably might be challenging, and even if it were accomplished it does not appear that the bound could be relaxed by more than an order of magnitude (still at least 3 orders of magnitude above the Z mass).

\section{\bf Acknowledgements}

The author would like to acknowledge useful conversations on this topic with Michael Dine, Thomas Banks, Howard Haber, Guido Festuccia, and David E. Kaplan.  The author would especially like to thank John Mason, who was involved in the early stages of this project, for many useful discussions, suggestions, and help.

%
%
%
%

\appendix

\section{Appendix: Miscellaneous Plots}
\label{appendix}

This appendix contains some plots involving large and small $\tan\beta$ and a large $\lambda$ coupling in front of a $SM\overline{M}$ term.

\begin{figure}[H]
\centering
\includegraphics[width=0.7\textwidth]{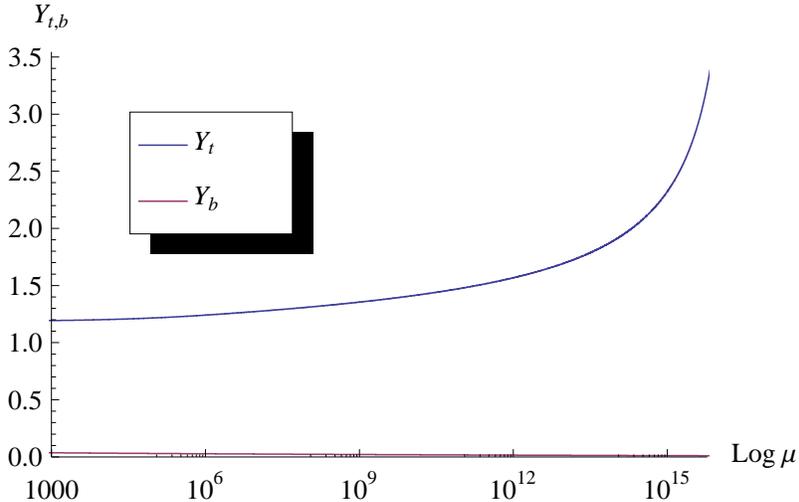}
\caption{A sample plot of the running of the top and bottom Yukawa couplings for small $\tan\beta$, where the top Yukawa reaches its perturbativity limit of $Y_t = \sqrt{4\pi}$ just before unification.  This tends to happen for small $\tan\beta$ if the SUSY scale is too low or if the extra flavors are too high.  The value of $\tan\beta$ at the electroweak scale used here is 1.35, however the precise value of $\tan\beta$ corresponding to this plot would be modified because the 1-loop corrections to the top and bottom masses and Yukawa couplings at the weak scale were not included in our analysis.  This subtlety should not affect our conclusions about the perturbativity of the gauge couplings themselves.}
\label{fig:yukawaplot01}
\end{figure}

\begin{figure}[H]
\centering
\includegraphics[width=1.1\textwidth]{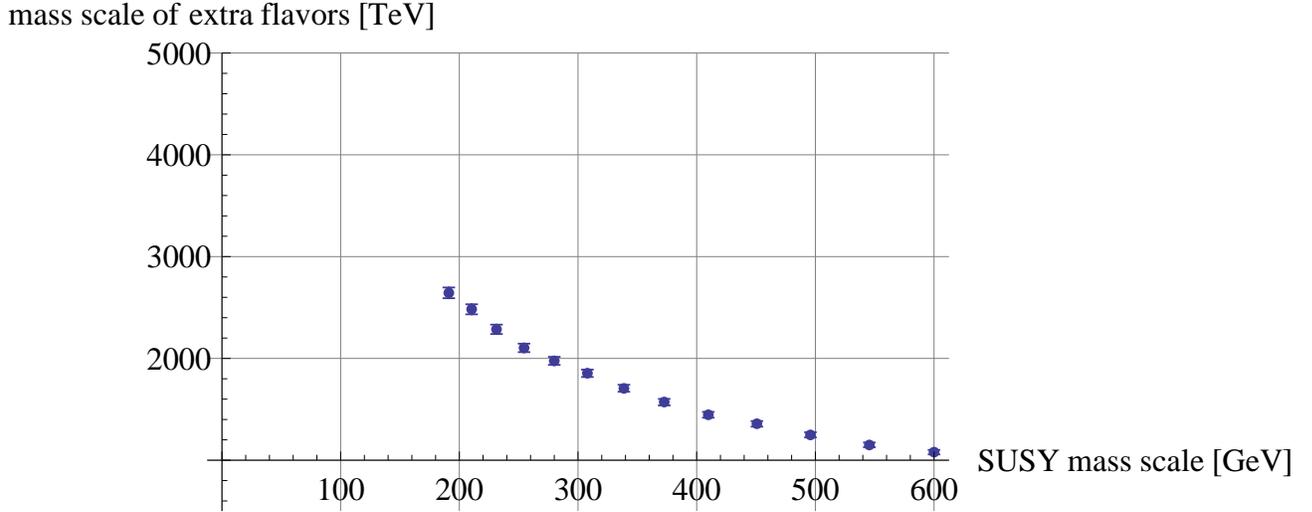}
\caption{A plot of the boundary below which 2-loop perturbativity is lost before unification occurs, with $\sin^2\hat{\theta}_W$ tuned at each point to minimize unification error, and $\tan\beta = 48$.  If the SUSY scale is below about 150~GeV, then the Yukawa couplings become non-perturbative, regardless of where the extra flavors are.  Above 150~GeV, the curve is not appreciably different from the $\tan\beta = 10$ plot in Figure~\ref{fig:threshplot01}.}
\label{fig:threshplot03}
\end{figure}

\begin{figure}[H]
\centering
\includegraphics[width=1.1\textwidth]{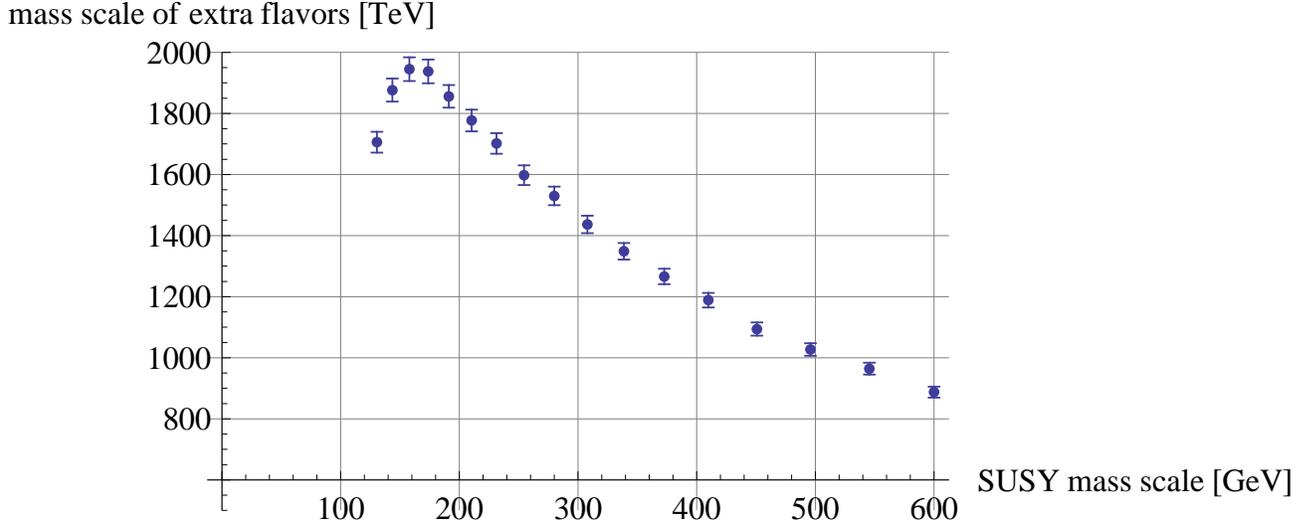}
\caption{A plot of the boundary below which 2-loop perturbativity is lost before unification occurs, with $\sin^2\hat{\theta}_W$ tuned at each point to minimize unification error, and $\tan\beta = 1.4$.  If the SUSY scale is below about 130~GeV, then the top Yukawa coupling becomes non-perturbative, regardless of where the extra flavors are.  Above 130~GeV, the shape of the curve is modified noticeably from the $\tan\beta = 10$ plot in Figure~\ref{fig:threshplot01}, and overall the boundary is a bit lower.}
\label{fig:threshplot04}
\end{figure}

\begin{figure}[H]
\centering
\includegraphics[width=0.7\textwidth]{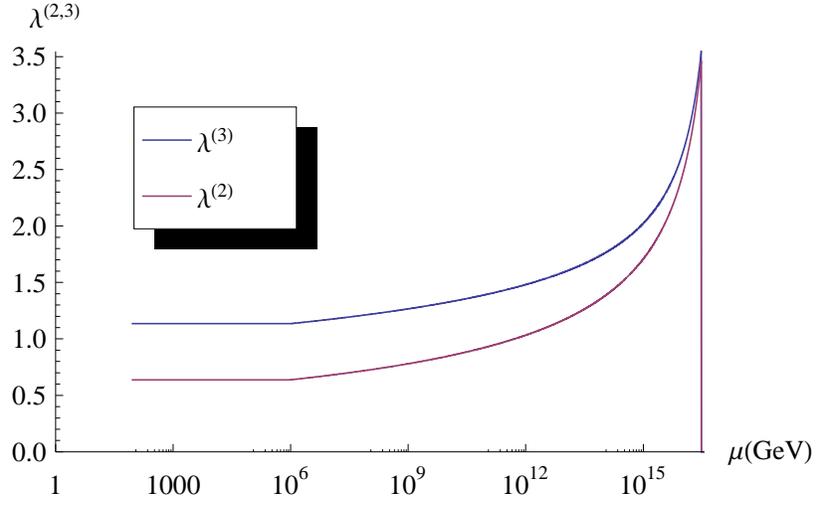}
\caption{A sample plot of the running of $\lambda^{(2)}$ and $\lambda^{(3)}$, where they unify at the GUT scale right at their perturbativity limit ($\lambda = \lambda^{(2)} = \lambda^{(3)} = \sqrt{4\pi}$).  Since these are couplings between the extra fundamentals and an extra singlet, they run only above the scale where the extra flavors enter.  Their values at the weak scale are $\lambda^{(2)} = 0.638$ and $\lambda^{(3)} = 1.135$}
\label{fig:lambdaplot01}
\end{figure}

\begin{figure}[H]
\centering
\includegraphics[width=1.1\textwidth]{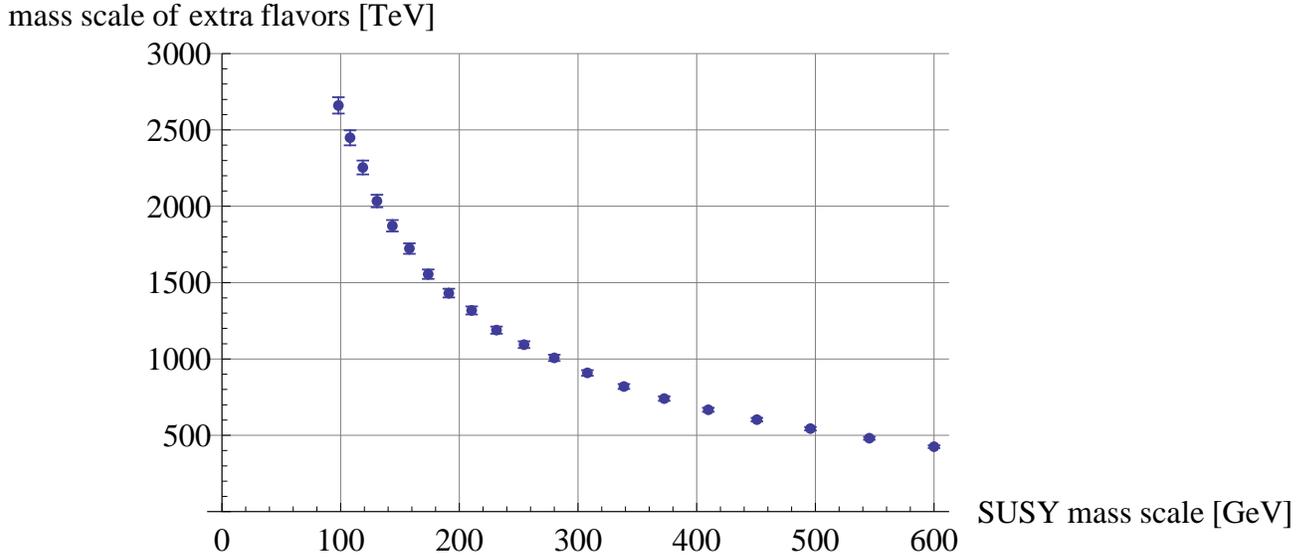}
\caption{A plot of the boundary below which 2-loop perturbativity is lost before unification occurs, with $\sin^2\hat{\theta}_W$ tuned at each point to minimize unification error, and boundary conditions at the weak scale of $\lambda^{(2)} = 0.638$, $\lambda^{(3)} = 1.118$.  The $\lambda$ couplings are about as strong as they can get without becoming non-perturbative for any of the plot range.  The shape of the curve is similar to when the $\lambda$ coupling is turned off, but shifted downward from the plot in Figure~\ref{fig:threshplot01} by roughly a factor of 2.  The rightmost point represents the lowest possible scale for the extra flavors of any of the scenarios run.}
\label{fig:threshplot08}
\end{figure}

\begin{figure}[H]
\centering
\includegraphics[width=1.1\textwidth]{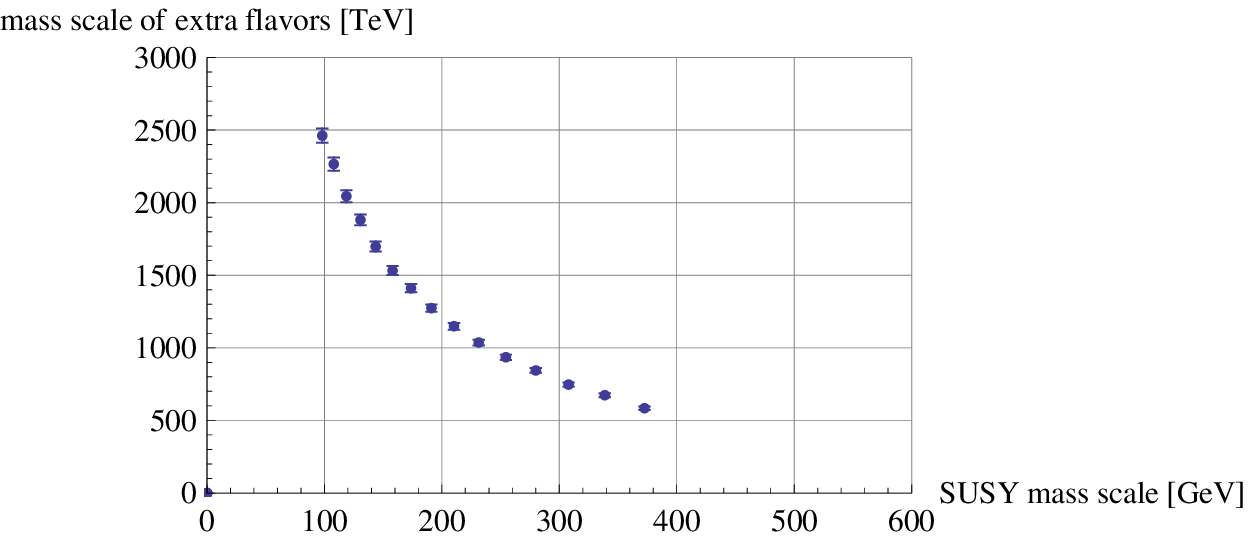}
\caption{A plot of the boundary below which 2-loop perturbativity is lost before unification occurs, with $\sin^2\hat{\theta}_W$ tuned at each point to minimize unification error, and boundary conditions at the weak scale of $\lambda^{(2)} = 0.650$, $\lambda^{(3)} = 1.120$.  When the SUSY scale is greater than about 375~GeV, the lambda couplings always become non-perturbative before unification.  When the SUSY scale is below 375~GeV, the $\lambda$ couplings are near their limit but stay perturbative.  The curve is shifted downward compared to Figure~\ref{fig:threshplot01} where the $\lambda$ coupling is turned off.}
\label{fig:threshplot09}
\end{figure}

\newpage

%
%
%
%
%
%

\end{document}